\documentclass[lettersize,journal]{IEEEtran}

\usepackage[table]{xcolor}

\usepackage{amsmath,amsfonts,amssymb,bm}

\usepackage{booktabs}
\usepackage{multirow}
\usepackage{tabu,longtable}
\usepackage{diagbox}
\usepackage{threeparttable}
\usepackage{makecell}
\usepackage{hhline}

\usepackage{graphicx}
\usepackage[caption=false,font=normalsize,labelfont=sf,textfont=sf]{subfig}

\usepackage{tikz}
\usepackage{pgfplots}
\pgfplotsset{compat=newest}

\usepackage{algorithm,algorithmic}

\usepackage{acronym}
\acrodef{otfs}[OTFS]{orthogonal time frequency space}
\acrodef{isl}[ISL]{inter-satellite link}
\acrodef{dmimo}[D-MIMO]{distributed multiple input multiple output}
\acrodef{pnt}[PNT]{positioning, navigation, and timing}
\acrodef{islac}[ISLAC]{integrated sensing, localization, and communications}
\acrodef{disac}[DISAC]{distributed integrated sensing and communications}
\acrodef{dislac}[DISLAC]{distributed ISLAC}
\acrodef{3gpp}[3GPP]{3rd generation partnership project}
\acrodef{leo}[LEO]{low Earth orbit}
\acrodef{meo}[MEO]{medium Earth orbit}
\acrodef{geo}[GEO]{geostationary Earth orbit}
\acrodef{ut}[UT]{user terminal}
\acrodef{upa}[UPA]{uniform planar array}
\acrodef{csi}[CSI]{channel state information}
\acrodef{ofdm}[OFDM]{orthogonal frequency division multiplexing}
\acrodef{los}[LOS]{line-of-sight}
\acrodef{toa}[TOA]{time-of-arrival}
\acrodef{pace}[PACE]{positioning-aided channel estimation}
\acrodef{mcrb}[MCRB]{misspecified Cramér-Rao bound}
\acrodef{awgn}[AWGN]{additive white Gaussian noise}
\acrodef{crb}[CRB]{Cramér-Rao bound}
\acrodef{cfo}[CFO]{carrier frequency offset}
\acrodef{peb}[PEB]{position error bound}
\acrodef{crb}[CRB]{Cramér-Rao bound}
\acrodef{lb}[LB]{lower bound}
\acrodef{rmse}[RMSE]{root mean squared error}
\acrodef{fim}[FIM]{Fisher information matrix}
\acrodef{tdd}[TDD]{time division duplex} 
\acrodef{wmmse}[WMMSE]{weighted minimal mean squared error} 
\acrodef{qcqp}[QCQP]{quadratically constrained quadratic program}
\acrodef{mrt}[MRT]{maximum ratio transmission}
\acrodef{gnss}[GNSS]{global navigation satellite system}
\acrodef{itu}[ITU]{International Telecommunication Union}
\acrodef{pab}[PAB]{Position-Aided Beamforming}
\acrodef{vdb}[VDB]{Vertically Directed Beamforming}
\acrodef{bs}[BS]{base station}
\acrodef{ntn}[NTN]{non-terrestrial networks}
\acrodef{uav}[UAV]{unmanned aerial vehicle}
\acrodef{tdma}[TDMA]{time-division multiple access}
\acrodef{haps}[HAPS]{high-altitude platform stations}
\acrodef{6g}[6G]{the sixth generation}
\acrodef{bse}[BSE]{beam squint effect}
\acrodef{cp}[CP]{cyclic prefix}
\acrodef{elaa}[ELAA]{extremely large antenna array}
\acrodef{ff}[FF]{far-field}
\acrodef{las}[L\&S]{localization and sensing}
\acrodef{nf}[NF]{near-field}
\acrodef{prs}[PRS]{positioning reference signal}
\acrodef{rf}[RF]{radio frequency}
\acrodef{ris}[RIS]{reconfigurable intelligent surface}
\acrodef{rtt}[RTT]{round-trip-time}
\acrodef{snr}[SNR]{signal-to-noise ratio}
\acrodef{sns}[SNS]{spatial non-stationarity}
\acrodef{swm}[SWM]{spherical wave model}
\acrodef{sar}[SAR]{synthetic aperture radar}
\acrodef{siso}[SISO]{single-input-single-output}
\acrodef{mimo}[MIMO]{multi-input-multi-output}
\acrodef{ue}[UE]{user equipment}
\acrodef{sp}[SP]{scatter point}
\acrodef{nlos}[NLOS]{non-line-of-sight}
\acrodef{tdoa}[TDOA]{time-difference-of-arrival}
\acrodef{am}[AM]{artificial multipath}
\acrodef{an}[AN]{artificial noise}
\acrodef{psd}[PSD]{power spectral density}
\acrodef{pdf}[PDF]{probability distribution function}
\acrodef{aoa}[AOA]{angle-of-arrival}
\acrodef{aod}[AOD]{angle-of-departure}
\acrodef{moo}[MOO]{multi-objective optimization}
\acrodef{qos}[QoS]{quality of service}
\acrodef{sdp}[SDP]{semi-definite programming}
\acrodef{lmi}[LMI]{linear matrix inequality}
\acrodef{sdr}[SDR]{semi-definite relaxation}
\acrodef{rcs}[RCS]{radar cross section}
\acrodef{isac}[ISAC]{integrated sensing and communication}
\acrodef{pdd}[PDD]{penalty dual decomposition}
\acrodef{bcd}[BCD]{block coordinate descent}

\usepackage{textcomp}
\usepackage{stfloats}
\usepackage{url}
\usepackage{verbatim}
\usepackage{cite}
\usepackage{units}

\begin{document}
\title{Distributed Integrated Sensing, Localization, and Communications over LEO Satellite Constellations}

\author{Yuchen Zhang, Francis Soualle, Musa Furkan Keskin, Yuan Liu, Linlong Wu, José A. del Peral-Rosado, Bhavani Shankar M. R., Gonzalo Seco-Granados, Henk Wymeersch, and Tareq Y. Al-Naffouri}

\markboth{draft}{draft}
\maketitle


\begin{abstract}
Low Earth orbit (LEO) satellite constellations are rapidly becoming essential enablers of next-generation wireless systems, offering global broadband access, high-precision localization, and reliable sensing beyond terrestrial coverage. However, the inherent limitations of individual LEO satellites, including restricted power, limited antenna aperture, and constrained onboard processing, hinder their ability to meet the growing demands of 6G applications. To address these challenges, this article introduces the concept of distributed integrated sensing, localization, and communication (DISLAC) over LEO constellations, inspired by distributed multiple input multiple output architectures. 
By enabling inter-satellite cooperation through inter-satellite links, DISLAC jointly exploits communication, localization, and sensing functionalities, achieving synergistic gains in throughput, positioning accuracy, and sensing robustness through shared resources and cooperative design.
We present illustrative case studies that quantify these benefits and analyze key system-level considerations, including synchronization, antenna reconfigurability, and inter-satellite link design. The article concludes by outlining open research directions to advance the practical deployment of DISLAC in future non-terrestrial networks.
\end{abstract}


\vspace{-5mm}

\section{Introduction} 
\Ac{leo} satellite systems are becoming integral to the 6G vision as a key enabler of emerging \ac{ntn}, offering broadband connectivity across land, sea, air, and space. Their ability to operate beyond the reach of terrestrial infrastructure helps bridge the digital divide and ensures resilient connectivity during disasters and in remote regions. The technical landscape is evolving accordingly. The \ac{3gpp} has standardized \ac{ntn} for integration into both enhanced mobile broadband and machine-type communications \cite{3gpp.38.811}. Beyond connectivity, \ac{leo} satellites now support diverse services including Earth observation, environmental sensing, and independent \ac{pnt}, complementing or substituting \ac{gnss} \cite{shicong2021leo,yin2024netmag,stock2025snt}. This evolution motivates extending the \ac{islac} paradigm \cite{henk2025disac}, which integrates sensing, localization, and communication, from terrestrial platforms to space-based architectures, enabling multifunctional \ac{leo} systems capable of operating far beyond ground network coverage.

However, individual \ac{leo} satellites remain limited by tight power budgets, small antenna apertures, and constrained onboard processing, making it difficult to meet the demands of high-throughput communication and precise sensing or localization. To overcome these limitations, we introduce the concept of \emph{\ac{dislac} over \ac{leo} constellations}. Drawing from the principles of \ac{dmimo} in terrestrial systems \cite{henk2025dmimo,henk2025disac}, \ac{dislac} leverages the scale and density of modern constellations and enables cooperation among satellites through \acp{isl}. Such cooperation enables flexible beamforming to boost communication rates and enhances localization and sensing via spatial diversity and fused low-\ac{snr} observations.
Existing studies on this topic can be broadly grouped into two threads: 
(i) \textit{single-satellite} \ac{islac} systems, where all functionalities are implemented on a single \ac{leo} satellite without inter-satellite cooperation~\cite{yin2024netmag,you2024twc}; and 
(ii) \textit{distributed} \ac{leo} constellations, where multiple satellites cooperate but treat communication, localization, and sensing as separate functions~\cite{halim2022oj,wolfgang2025twc}. 
Among these,~\cite{wolfgang2025twc} jointly considers communication and sensing, yet omits \ac{ue} localization and overlooks key system-level aspects such as synchronization, \ac{isl} design, and network topology. 
Hence, a unified framework that jointly integrates all three functionalities across distributed \ac{leo} constellations and captures their mutual interactions remains unexplored.

\begin{figure*}[t]
  \centering
  \includegraphics[width=\linewidth]{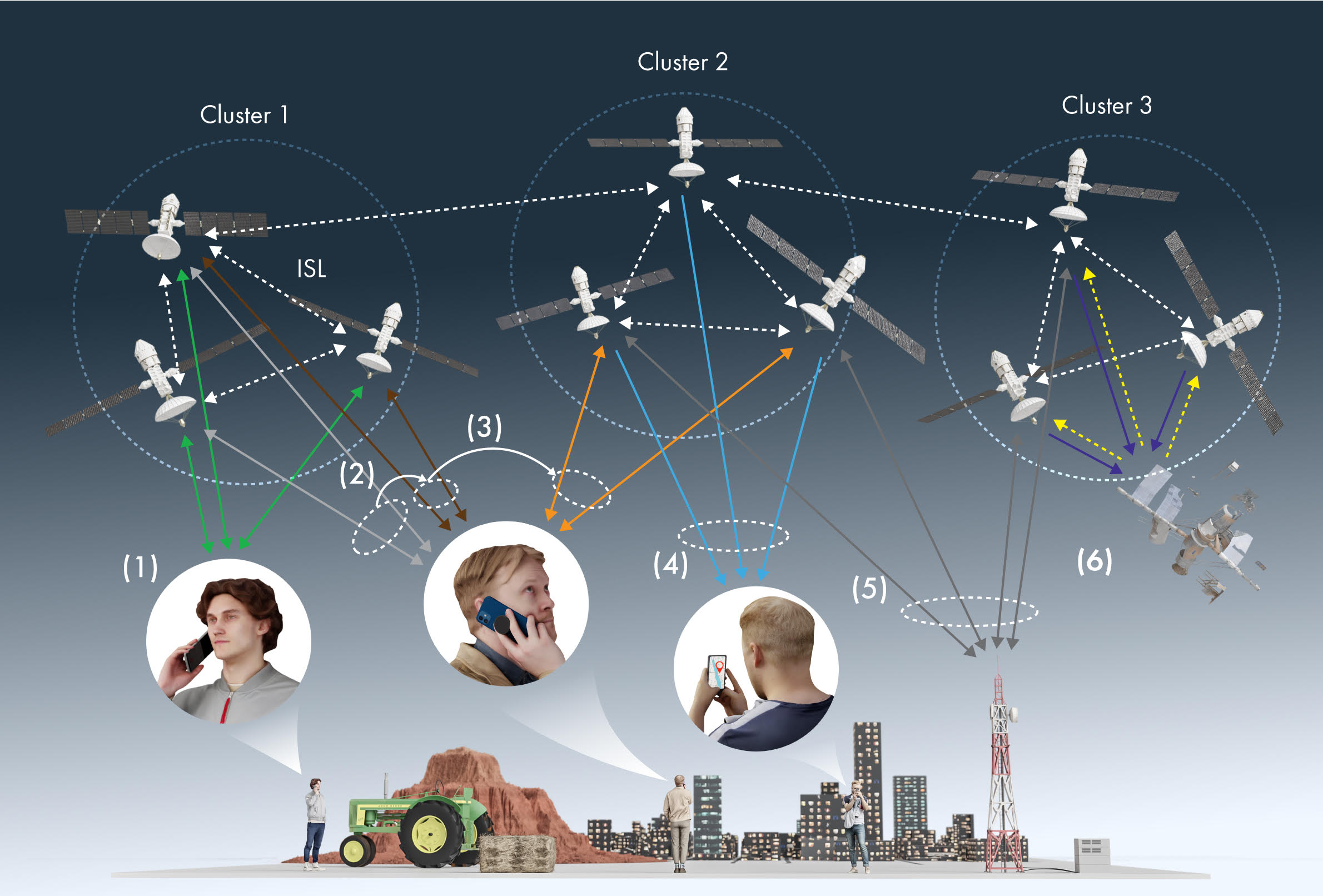}
  \vspace{-1.8em}
  \caption{Overview of \ac{dislac} over \ac{leo} satellite constellations: (1) Multi-\ac{leo} cooperative beamforming; (2) Intra-cluster handover; (3) Inter-cluster handover; (4) Coordinated multi-\ac{leo} localization; (5) Multi-\ac{leo} cooperative backhauling; (6) Multi-static space debris sensing. }
  \label{sys_mod}
\end{figure*}

To the best of our knowledge, this article is the first to present a comprehensive \ac{dislac} framework that unifies communication, localization, and sensing at the constellation level, 
building on recent advances in \ac{disac}/\ac{dislac} for terrestrial \ac{dmimo} networks~\cite{henk2025dmimo,henk2025disac} and extending them to the \acp{ntn}.
As shown in Fig. \ref{sys_mod}, we consider a cooperative scenario in which multiple multi-antenna \ac{leo} satellites jointly perform communication, localization, and sensing with support of \acp{isl}. We analyze the benefits of this architecture through illustrative case studies and explore critical system-level considerations such as synchronization, ISL constraints, antenna flexibility, and scalability. Finally, we outline open research directions to guide the practical development of next-generation \ac{dislac}-enabled \acp{ntn}.

\section{Benefits of LEO constellation-level communication, localization, and sensing} 

In this section, we provide an overview of the usage of \ac{leo} satellite constellations for communication, localization and sensing, considering each function separately. This motivates the vision of a \ac{dislac} system that can support the three functions jointly, as elaborated in Section IV.

\subsection{Communication Perspective} 
Historically, \ac{leo} satellite systems have been pivotal in extending connectivity to remote and underserved regions due to their lower orbital altitudes (typically 500–1200 km), which significantly reduce signal propagation delays compared to \ac{geo} satellites. While early deployments like Iridium and Globalstar were constrained by high deployment costs, sparse constellations, and basic onboard capabilities, next-generation systems such as Starlink, OneWeb, and Kuiper now leverage high-gain phased-array antennas, efficient \ac{rf} chains, and dynamic beamforming, achieving data rates up to hundreds of Mbps to support high-demand applications (e.g. HD video streaming). However, these systems typically employ single-satellite-to-user access, which remains vulnerable to outages due to terrain blockage or severe weather, limits user experience due to constrained satellite power and antenna size, and suffers from frequent handovers caused by high \ac{leo} satellite mobility and short visibility durations. These issues lead to inconsistent user experiences and inefficient resource utilization, challenges further amplified by the demanding requirements of emerging 6G applications, including extended reality (XR) and large-scale data-driven edge computing.

These challenges motivate the paradigm shift toward networked \ac{leo} satellite cooperative service, where satellites no longer operate independently but instead collaborate as part of a distributed spaceborne network, a concept that falls under the ambition of \ac{dislac} over \ac{leo} satellite constellations. In this architecture, multiple \ac{leo} satellites simultaneously serve the same ground terminal using collaborative beamforming, enabling better management of inter-satellite interference, enhanced link budget and robustness through macro diversity, and reduced frequency of inter-cluster handovers due to extended coverage. This is particularly valuable in high-throughput scenarios, where a single satellite is insufficient, or in adverse propagation environments, such as mountainous terrain or deep urban canyons, where the signal from one satellite may be obstructed while others remain visible. Meanwhile, inter-satellite data routing via \acp{isl} allows the networks to dynamically balance traffic across spatially distributed nodes. For instance, when one satellite experiences a high user load, neighboring satellites can offload part of the traffic through \acp{isl}, reducing latency and alleviating local congestion. Likewise, ground terminals in motion, such as those in vehicles, high-speed trains, or aircraft, can be seamlessly handed over between satellites without disrupting connectivity, as neighboring satellites jointly track/serve the users across their overlapping coverage footprints.

\subsection{Localization Perspective} 
\ac{leo} satellite constellations offer a promising platform for positioning, complementing or even substituting \ac{gnss}, especially in \ac{gnss}-challenged environments. Their lower altitude results in stronger signal reception and improved geometric diversity compared to \ac{gnss}. When combined with coordinated waveform design and inter-satellite cooperation, \ac{leo} satellite-based localization provides improved accuracy and robustness, forming a scalable and resilient complement to both terrestrial and \ac{gnss} systems, paving the way for global high-accuracy positioning in 6G networks \cite{Saleh2025Integrated}.

The distributed nature of \ac{leo} satellites enables coordinated multi-satellite localization. In such setups, multiple satellites simultaneously transmit orthogonal waveforms (e.g., separated in frequency, code, or time) while remaining tightly synchronized in time and frequency. The \ac{ue} can jointly estimate delay, Doppler, or angle-based observables from these signals. Using \ac{ofdm} as waveform candidate is natural due to its compatibility with communications and favorable time-frequency resolution, though high Doppler and delay spreads in \ac{leo} scenarios must be managed carefully. Techniques such as timing advance and Doppler pre-compensation at the satellite help mitigate these effects.

While single-\ac{leo} satellite localization is theoretically feasible by exploiting the satellite's motion to form a synthetic aperture over time, this method is limited to scenarios with static or slowly moving \acp{ue} and requires long observation intervals, high signal stability, and favorable trajectories. These constraints limit its practicality, particularly in latency-sensitive or mobile applications. Multi-\ac{leo} satellite localization, in contrast, offers geometric diversity, reduced reliance on \ac{ue} motion, and faster, more accurate positioning.
Additional benefits arise when \ac{leo} satellites are also phase-synchronized. This enables the use of carrier-phase measurements for high-precision positioning, akin to real-time kinematic (RTK) or precise point positioning (PPP) in \ac{gnss}. Moreover, coherent joint transmission forms a virtual distributed array, and creates near-field effects and wavefront curvature, which can significantly enhance localization accuracy, potentially to sub-wavelength levels. Practical schemes include spatial beamforming among multiple anchors (for high SNR, but limited coverage) or orthogonal transmission (lower SNR, but wide coverage).

\subsection{Sensing Perspective} 
\ac{leo} satellite-based sensing plays a critical role in space situational awareness, enabling important tasks such as debris detection, collision warnings, and identification of orbiting objects~\cite{yuan2024icassp}. Compared to ground-based methods like terrestrial radars and optical telescopes, satellite-based sensing provides broader spatial coverage and operates independently from ground-level constraints, such as adverse weather conditions.

Radar-based sensing systems operating at radio frequencies have been successfully deployed in space for decades, with notable functionalities including TerraSAR-X and RadarSat-2. These systems utilize \ac{sar} technology, enabling detailed imaging and effective detection and object characterization. However, single-satellite radar approaches suffer inherent constraints especially when dealing with space moving target like debris or other hostile moving objects. Limited viewing angles and infrequent observation windows restrict their accuracy in localization and parameter estimation, making the opportunistic sensing highly unreliable in space environment and undermining its practical application. Moreover, reliance on a single satellite introduces vulnerability, and any technical failures or obstructions can severely compromise the system's effectiveness.

Recently, the emergence of \ac{leo} satellite constellations offers a compelling solution to these limitations, with Starlink being a renowned example. Unlike traditional single-satellite monostatic radars, distributed multi-\ac{leo} satellite sensing naturally supports bistatic and multistatic radar operations, significantly enriching the quality of sensed information by exploiting spatial diversity. Observations of a target from different geometric angles improve both the detection performance and parameter estimation accuracy. For instance, the Doppler velocity of space debris can be more accurately and rapidly determined when observed simultaneously by at least two \ac{leo} satellites from distinct orbital positions, facilitating timely collision-avoidance maneuvers. Furthermore, multi-\ac{leo} satellite sensing circumvents to some extent the limitations of power consumption, hardware complexity, and cost associated with deploying large antenna arrays on individual satellites. Leveraging collaborative beamforming, multiple \ac{leo} satellites can form focused beams directed precisely toward targets, achieving improved detection sensitivity and angular resolution while substantially reducing the power and antenna requirements for each satellite individually. Such an approach not only improves radar performance but also offers practical scalability compared to single-satellite radar solutions. Finally, the presence of \acp{isl} within networks naturally increases opportunities for opportunistic sensing, enabling continuous or near-continuous monitoring of targets.

\subsection{Mutual Benefits and Trade-offs}
\begin{figure}[t] 
		\centering
		\includegraphics[width=1.0 \linewidth]{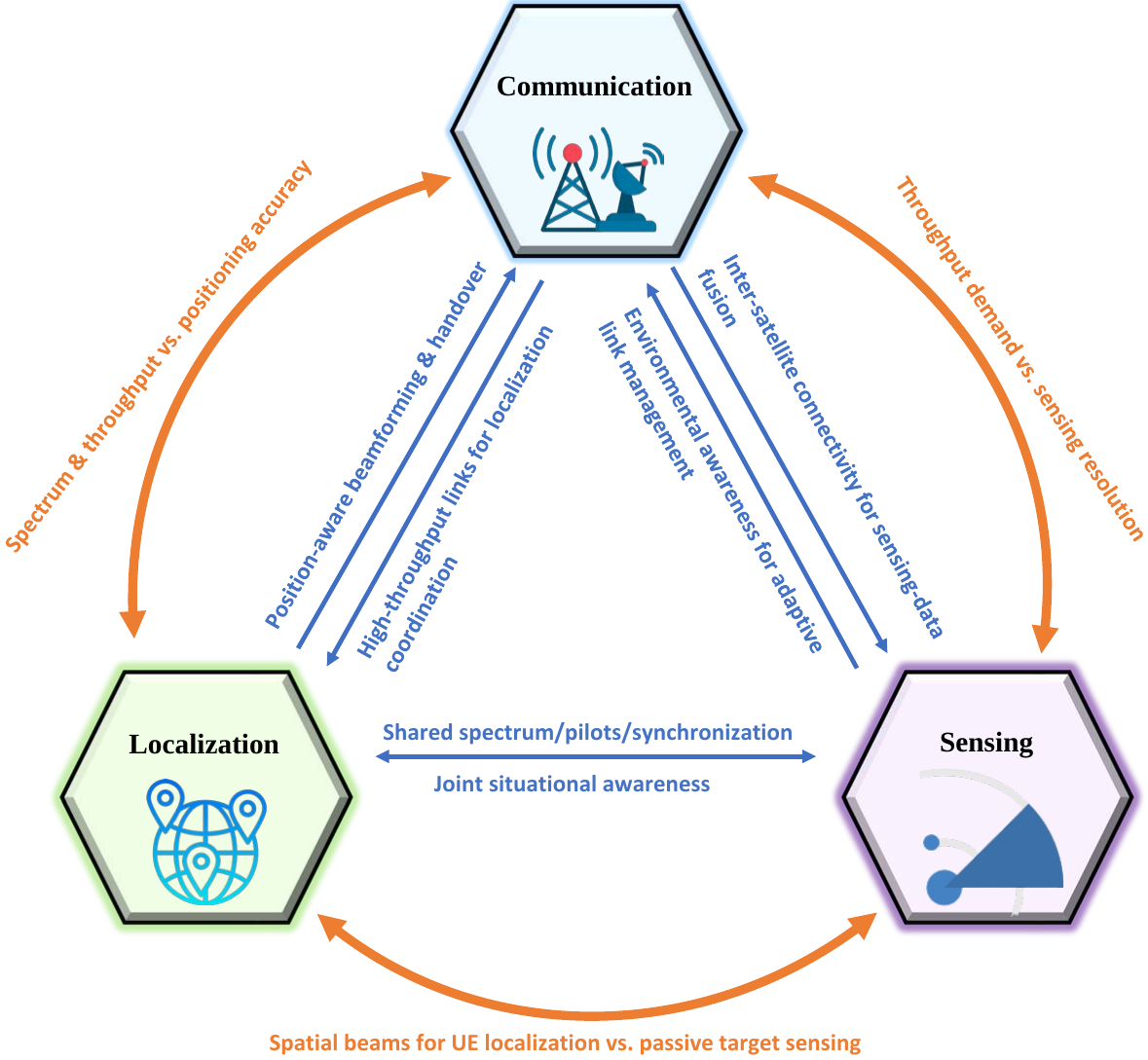}
		\caption{Conceptual illustration of exemplary synergies and trade-offs among communication, localization, and sensing in a unified \ac{dislac} framework over \ac{leo} constellations. “Communication” aggregates service links between \acp{leo} and \acp{ue} and \acp{isl}; “Localization” covers \ac{ue} positioning and network-wide satellite localization; “Sensing” includes ground and inter-\ac{leo} sensing. Blue arrows illustrate example mutual benefits, such as position-aware beamforming and handover, high-throughput links for localization coordination, \ac{isl}-enabled fusion of sensing data, and shared spectrum/pilots/synchronization for joint situational awareness. Orange arcs indicate representative resource trade-offs in spectrum/throughput versus positioning accuracy, throughput demand versus sensing resolution, and spatial beam allocation between \ac{ue} localization and passive target sensing.}
		\label{fig:dislac_tradeoffs}
\end{figure}

In a unified \ac{dislac} framework, communication, localization, and sensing become mutually reinforcing rather than isolated functionalities, as conceptually illustrated in Fig.~\ref{fig:dislac_tradeoffs}. Accurate \ac{ue} positioning and network-wide \ac{leo} localization can significantly improve geometry-dependent channel estimation, enable position-aware beamforming, and support efficient user association, scheduling, and handover management. Conversely, high-throughput service links and \acp{isl} provide the data backbone for distributing sensing observations and localization coordination across the constellation, as well as for fusing multi-\ac{leo} sensing data. Sensing measurements, in turn, enhance situational awareness by detecting obstacles, atmospheric disturbances, or debris, thereby supporting adaptive link management and robust service continuity. These synergies also introduce inherent resource trade-offs: spectrum and waveform resources must be shared between throughput and positioning accuracy, sensing resolution competes with communication throughput, and spatial beam patterns need to be allocated between illuminating \acp{ue} for localization and illuminating passive targets for sensing. Balancing these coupled benefits and trade-offs calls for flexible resource orchestration and adaptive \ac{isl} coordination, highlighting the distinctive system-level design challenges of integrating communication, localization, and sensing in distributed \ac{leo} constellations.

\section{Case Studied in LEO constellation-level communication, localization, and sensing} 
Case studies that quantitatively evaluate the benefits and challenges of \ac{dislac} over \ac{leo} satellite constellations are presented, with separate analyses for communication, localization, and sensing to reveal the distinct implications of distributed cooperation for each function.

\begin{figure}[t]
\centering
\begin{minipage}[b]{0.49\linewidth}
  \centering
%
%
\definecolor{mycolor1}{rgb}{0.00000,0.44700,0.74100}%
\definecolor{mycolor2}{rgb}{0.85000,0.32500,0.09800}%
\definecolor{mycolor3}{rgb}{0.92900,0.69400,0.12500}%
\definecolor{mycolor4}{rgb}{0.49400,0.18400,0.55600}%
\definecolor{mycolor5}{rgb}{0.46600,0.67400,0.18800}%
\definecolor{mycolor6}{rgb}{0.30100,0.74500,0.93300}%
\definecolor{mycolor7}{rgb}{0.63500,0.07800,0.18400}%
\begin{tikzpicture}

\begin{axis}[%
width=33mm,
height=44mm,
scale only axis,
xtick={4, 8, 16, 32}, 
xmin=4,
xmax=32,
xlabel style={font=\color{white!15!black}, font=\footnotesize, yshift=6pt},
xlabel={LEO Satellite Number $L$},
ymin=0,
ymax=700000000,
ylabel style={font=\color{white!15!black}, font=\footnotesize, yshift=-5pt, xshift=0pt},
ylabel={Sum Rate [bps]},
axis background/.style={fill=white},
xmajorgrids,
ymajorgrids,
legend style={at={(0.000609,0.604)}, anchor=south west, legend cell align=left, align=left, draw=white!15!black, font=\footnotesize, legend columns = 1}
]

\addplot [color=mycolor3, line width=2.0pt, mark size=3.0pt, mark=asterisk, mark options={solid, mycolor3}]
  table[row sep=crcr]{%
4	27434048.9976905\\
8	91522293.0675284\\
16	253740534.233744\\
32	625909786.900351\\
};
\addlegendentry{Centralized}

\addplot [color=mycolor1, line width=2.0pt, mark size=3.0pt, mark=+, mark options={solid, mycolor1}]
  table[row sep=crcr]{%
4	26748224.2249015\\
8	88619865.8750947\\
16	250754915.059286\\
32	606638351.748207\\
};
\addlegendentry{Ring}

\addplot [color=mycolor2, line width=2.0pt, mark size=3.0pt, mark=o, mark options={solid, mycolor2}]
  table[row sep=crcr]{%
4	26187017.9065153\\
8	80078367.4679532\\
16	226586903.137003\\
32	580799140.791709\\
};
\addlegendentry{Star}



\addplot [color=mycolor6, dashed, line width=2.0pt, mark size=3.0pt, mark=asterisk, mark options={solid, mycolor6}]
  table[row sep=crcr]{%
4	5555930.98380753\\
8	7417060.56674254\\
16	9494929.29058925\\
32	17120813.8460533\\
};
\addlegendentry{$\text{S}^3$}



\end{axis}
\end{tikzpicture}%
    \vspace{-1.cm}
  \centerline{(a)} \medskip
\end{minipage}
\begin{minipage}[b]{0.49\linewidth}
  \centering
%
%
\definecolor{mycolor1}{rgb}{0.00000,0.44700,0.74100}%
\definecolor{mycolor2}{rgb}{0.85000,0.32500,0.09800}%
\definecolor{mycolor3}{rgb}{0.92900,0.69400,0.12500}%
\definecolor{mycolor4}{rgb}{0.49400,0.18400,0.55600}%
\definecolor{mycolor5}{rgb}{0.46600,0.67400,0.18800}%
\begin{tikzpicture}

\begin{axis}[%
width=33mm,
height=44mm,
at={(0mm, 0mm)},
scale only axis,
xtick={4, 8, 16, 32}, 
xmin=4,
xmax=32,
xlabel style={font=\color{white!15!black}, font=\footnotesize, yshift=6pt},
xlabel={UE Number $U$ },
ymin=0,
ymax=18000000,
ylabel style={font=\color{white!15!black}, font=\footnotesize, yshift=-5pt, xshift=0pt},
ylabel={Overhead Number},
axis background/.style={fill=white},
xmajorgrids,
ymajorgrids,
legend style={font=\footnotesize, at={(0.000609,0.528)}, anchor=south west, legend cell align=left, align=left, draw=white!15!black}
]
\addplot [color=mycolor5 , line width=2.0pt, mark size=3.0pt, mark=o, mark options={solid, mycolor5}]
  table[row sep=crcr]{%
4	368640\\
8	1228800\\
16	4423680\\
32	16711680\\
};
\addlegendentry{Star (C, $L = 16$)}

\addplot [color=mycolor4 , line width=2.0pt, mark size=3.0pt, mark=o, mark options={solid, mycolor4}]
  table[row sep=crcr]{%
4	172032\\
8	573440\\
16	2064384\\
32	7798784\\
};
\addlegendentry{Star (C, $L = 8$)}

\addplot [color=mycolor3 , line width=2.0pt, mark size=3.0pt, mark=o, mark options={solid, mycolor3}]
  table[row sep=crcr]{%
4	73728\\
8	245760\\
16	884736\\
32	3342336\\
};
\addlegendentry{Star (C, $L = 4$)}

\addplot [color=mycolor2, line width=2.0pt, mark size=3.0pt, mark=o, mark options={solid, mycolor2}]
  table[row sep=crcr]{%
4	24576\\
8	81920\\
16	294912\\
32	1114112\\
};
\addlegendentry{Star (E)}

\addplot [color=mycolor1, line width=2.0pt, mark size=3.0pt, mark=+, mark options={solid, mycolor1}]
  table[row sep=crcr]{%
4	24576\\
8	81920\\
16	294912\\
32	1114112\\
};
\addlegendentry{Ring}

\end{axis}
\end{tikzpicture}%
    \vspace{-1.cm}
  \centerline{(b)} \medskip
\end{minipage}
\vspace{-0.8cm}
\caption{Throughput and signaling overhead comparison: (a) Sum-rate versus number of \ac{leo} satellites $L$ under different beamforming schemes; (b) Signaling overhead of collaborative beamforming versus number of UEs $U$, evaluated under Ring and Star topologies with varying satellite number $L$. (E: Edge node, C: Central node).}
\label{dist_bf_compare}
\vspace{-0.2cm}
\end{figure}
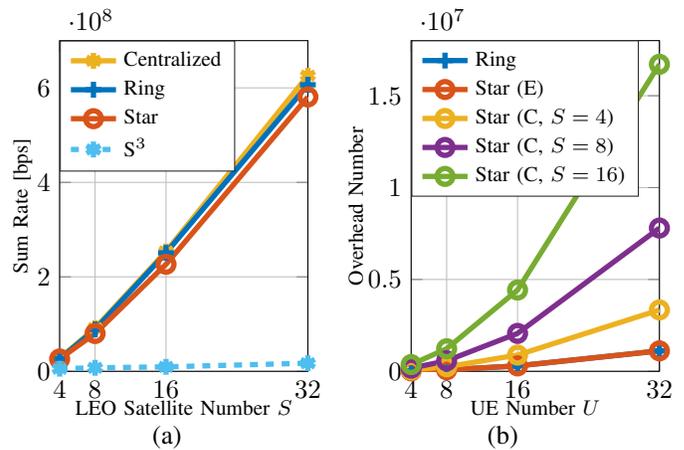

\subsection{Case Study 1: Boosting Communication Throughput via Multi-\ac{leo} Satellite Collaborative Beamforming}
In this case study, we consider a scenario where time-synchronized \ac{leo} satellites, each equipped with hybrid antenna arrays, collaboratively serve multiple terrestrial \acp{ue} in the downlink via collaborative beamforming. Rather than optimizing the instantaneous sum rate, highly dependent on instantaneous \ac{csi} that is difficult to acquire due to severe Doppler shifts and latency, we evaluate the ergodic sum rate based on long-term channel statistics, which can be efficiently approximated using the closed-form hardening bound. For implementation details, we refer interested readers to our previous work \cite{Zack2025Scalable}.

Figures \ref{dist_bf_compare}(a) and (b) illustrate the sum rate and signaling overhead under different beamforming strategies, evaluated across varying numbers of satellites $L$ and users $U$. We adopt a decentralized \ac{wmmse} algorithm for collaborative beamforming under two typical \ac{leo} network topologies: Ring and Star. Unlike centralized approaches, where a central processing unit (CPU) jointly optimizes all beamformers using global \ac{csi}, the decentralized schemes perform local optimization at each satellite, relying only on the exchange of intermediate variables with neighboring satellites according to the respective topology \cite{Zack2025Scalable}. As a baseline, we also consider the single-satellite service ($\text{S}^3$) model, where each user is served by only one satellite.

As shown in Fig. \ref{dist_bf_compare}(a), both Ring and Star-based collaborative beamforming significantly outperform S\textsuperscript{3}, achieving near-centralized sum rate performance. However, decentralized operation entails cross-satellite signaling overhead due to iterative information exchange. Fig. \ref{dist_bf_compare}(b) shows that this overhead grows with the number of users. Notably, unlike Ring topology with balanced burden, the Star topology imposes an unbalanced burden on the central node, whose overhead increases with the number of satellites, while edge nodes experience constant overhead. Despite its balanced nature, the Ring topology typically suffers from longer latency due to its sequential information flow. These results underscore the need to carefully consider architectural and signaling constraints when implementing collaborative beamforming over \ac{leo} satellite constellations.

\subsection{Case Study 2: Delay and Doppler Characterization for Multi-\ac{leo} Satellite Positioning}
\label{ssec:CS2}
To support high-accuracy positioning, \ac{leo} satellites can be used to generate \ac{prs}. However, their movement and wide geometric spread also introduces significant Doppler shifts and varying propagation delays, which may challenge the design of reference signals and the associated receiver processing. To better understand these effects, we analyze the delay and Doppler characteristics of downlink signals from a dense set of overhead \ac{leo} satellites to a static user on the ground. The satellites are assumed to broadcast \ac{prs}-like waveforms toward Earth, and the user passively receives these signals for positioning purposes. While in practice only a few satellites would transmit \ac{prs} simultaneously, we evaluate a dense configuration to capture the full range of possible delays and Doppler shifts, thus providing a worst-case scenario for system design.

\begin{figure}
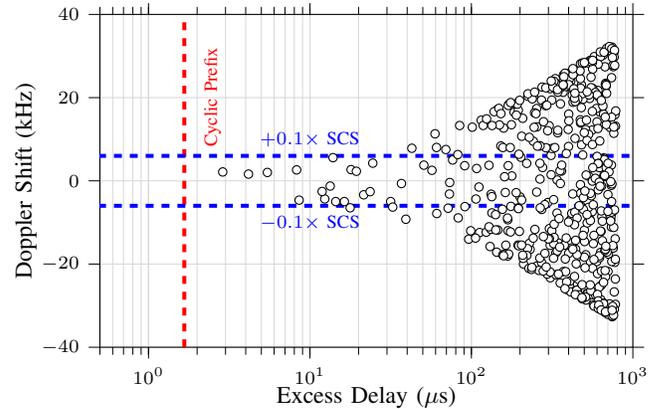

    \centering
    \include{Figures/Spreads}
    \vspace{-10mm}
\caption{Joint distribution of downlink delay and Doppler shift from 200 randomly placed \ac{leo} satellites to a static \ac{ue} on Earth. Each circle represents a satellite. The satellites are uniformly distributed above the \ac{ue} within zenith angles from $0^\circ$ to $5^\circ$ (i.e., elevation angles above $85^\circ$), at an altitude of 600~km and speed of 7.5~km/s. Satellites move at equal speed along randomly directed tangents to the orbital sphere. The carrier frequency is 2~GHz, and the \ac{ofdm} waveform uses a subcarrier spacing (SCS) of 60~kHz with a cyclic prefix of 1.6~$\mu$s. The vertical dashed red line marks the cyclic prefix duration; the horizontal dashed blue lines indicates $0.1 \times$~SCS, a typical threshold for acceptable Doppler spread.}
    \label{fig:spreads}    
\end{figure}

Figure \ref{fig:spreads} illustrates the joint distribution of delays and Doppler shifts as seen at the user terminal. Delay is governed by the distance between each satellite and the user (the shortest delay is subtracted), while Doppler depends on the projection of satellite motion along the line of sight. Overlaid threshold lines represent typical tolerances derived from \ac{ofdm}-based systems, such as the cyclic prefix and subcarrier spacing, which are relevant when reusing communication waveforms for positioning.
The results indicate that many satellites introduce severe delay and Doppler values, which leads to signal misalignment or reduced measurement accuracy. This suggests that significant time and frequency pre-compensation will be required when multiple \ac{leo} satellites are used jointly for positioning. Alternatively, the selection of \ac{prs}-transmitting satellites could be constrained based on real-time delay and Doppler estimates, improving robustness without requiring tight waveform constraints.

\subsection{Case Study 3: OFDM Waveform Exploration for Multi-\ac{leo} Satellite Multistatic Sensing}
We explore \ac{ofdm} waveforms for spaceborne sensing due to their inherent support for delay and Doppler processing and seamless compatibility with modern communication systems. However, applying \ac{ofdm} in \ac{leo} scenarios presents unique challenges, as satellite-based sensing must account for extended propagation distances (up to 500 km) and high relative velocities (up to 7.5 km/s) caused by orbital dynamics. Critical radar performance metrics, including maximum unambiguous range, range resolution, maximum unambiguous velocity, and velocity resolution, are highly sensitive to waveform configurations, particularly subband spacing.

\begin{figure}[t]
\centering
\begin{minipage}[b]{0.49\linewidth}
  \centering
    \definecolor{mycolor1}{rgb}{0.00000,0.00000,0.51562}%
\definecolor{mycolor2}{rgb}{0.00000,0.00000,0.95312}%
\definecolor{mycolor3}{rgb}{0.00000,0.40625,1.00000}%
\definecolor{mycolor4}{rgb}{0.00000,0.84375,1.00000}%
\definecolor{mycolor5}{rgb}{0.28125,1.00000,0.71875}%
\definecolor{mycolor6}{rgb}{0.73438,1.00000,0.26562}%
\definecolor{mycolor7}{rgb}{1.00000,0.82812,0.00000}%
\definecolor{mycolor8}{rgb}{1.00000,0.39062,0.00000}%
\definecolor{mycolor9}{rgb}{0.93750,0.00000,0.00000}%

\begin{tikzpicture}

\begin{axis}[%
    width=3.5cm,
    height=5.4cm,
    at={(0cm,0cm)},
    scale only axis,
    point meta min=2,
    point meta max=200,
    clip=false,
    xmin=-0.9522,
    xmax=1.0478,
    ymin=-0.5611,
    ymax=1.2389,
    axis line style={draw=none},
    ticks=none,
    axis x line*=bottom,
    axis y line*=left,
    xlabel style={font=\footnotesize},
    ylabel style={font=\footnotesize},
    tick label style={font=\scriptsize},
    legend style={font=\scriptsize, cells={anchor=west}},
    legend pos=north east
]

\addplot[area legend, draw=mycolor1, fill=mycolor1, fill opacity=0.3, forget plot]
table[row sep=crcr]{
x	y\\
0	0\\
-0.866	-0.5\\
0.866	-0.5\\
0	0\\
}--cycle;

\addplot[area legend, draw=mycolor2, fill=mycolor2, fill opacity=0.3, forget plot]
table[row sep=crcr]{
x	y\\
0	0.1111\\
-0.7698	-0.4444\\
0.7698	-0.4444\\
0	0.1111\\
}--cycle;

\addplot[area legend, draw=mycolor3, fill=mycolor3, fill opacity=0.3, forget plot]
table[row sep=crcr]{
x	y\\
0	0.2222\\
-0.6736	-0.3889\\
0.6736	-0.3889\\
0	0.2222\\
}--cycle;

\addplot[area legend, draw=mycolor4, fill=mycolor4, fill opacity=0.3, forget plot]
table[row sep=crcr]{
x	y\\
0	0.3333\\
-0.5774	-0.3333\\
0.5774	-0.3333\\
0	0.3333\\
}--cycle;

\addplot[area legend, draw=mycolor5, fill=mycolor5, fill opacity=0.3, forget plot]
table[row sep=crcr]{
x	y\\
0	0.4444\\
-0.4811	-0.2778\\
0.4811	-0.2778\\
0	0.4444\\
}--cycle;

\addplot[area legend, draw=mycolor6, fill=mycolor6, fill opacity=0.3, forget plot]
table[row sep=crcr]{
x	y\\
0	0.5556\\
-0.3849	-0.2222\\
0.3849	-0.2222\\
0	0.5556\\
}--cycle;

\addplot[area legend, draw=mycolor7, fill=mycolor7, fill opacity=0.3, forget plot]
table[row sep=crcr]{
x	y\\
0	0.6667\\
-0.2887	-0.1667\\
0.2887	-0.1667\\
0	0.6667\\
}--cycle;

\addplot[area legend, draw=mycolor8, fill=mycolor8, fill opacity=0.3, forget plot]
table[row sep=crcr]{
x	y\\
0	0.7778\\
-0.1925	-0.1111\\
0.1925	-0.1111\\
0	0.7778\\
}--cycle;

\addplot[area legend, draw=mycolor9, fill=mycolor9, fill opacity=0.3, forget plot]
table[row sep=crcr]{
x	y\\
0	0.8889\\
-0.0962	-0.0556\\
0.0962	-0.0556\\
0	0.8889\\
}--cycle;

\addplot[area legend, draw=black!50!red, fill=black!50!red, fill opacity=0.3, forget plot]
table[row sep=crcr]{
x	y\\
0	1\\
0	0\\
0	0\\
0	1\\
}--cycle;

\addplot [color=black, dashed, line width=1pt, forget plot]
table[row sep=crcr]{0 0\\ 0 1.1\\};
\addplot [color=black, dashed, line width=1pt, forget plot]
table[row sep=crcr]{0 0\\ -0.9526 -0.55\\};
\addplot [color=black, dashed, line width=1pt, forget plot]
table[row sep=crcr]{0 0\\ 0.9526 -0.55\\};

\node[font=\footnotesize] at (axis cs:-0.004,0.074) {0.08};
\node[font=\footnotesize] at (axis cs:-0.2,-0.084) {0.001};
\node[font=\footnotesize] at (axis cs:0.172,-0.084) {0.7};
\node[font=\footnotesize] at (axis cs:0,0.35) {0.44};
\node[font=\footnotesize] at (axis cs:-0.337,-0.199) {0.004};
\node[font=\footnotesize] at (axis cs:0.324,-0.199) {4.4};
\node[font=\footnotesize] at (axis cs:0,0.695) {2.56};
\node[font=\footnotesize] at (axis cs:-0.606,-0.35) {0.025};
\node[font=\footnotesize] at (axis cs:0.606,-0.35) {25.6};
\node[font=\footnotesize] at (axis cs:0,1.05) {15.00};
\node[font=\footnotesize] at (axis cs:-0.869,-0.52) {0.146};
\node[font=\footnotesize] at (axis cs:0.899,-0.525) {150.0};
\node[font=\footnotesize] at (axis cs:-0.052,1.204) {$v_{\text{max}}\text{ (km/s)}$};
\node[font=\footnotesize] at (axis cs:-0.828,-0.638) {$\Delta R\text{ (km)}$};
\node[font=\footnotesize] at (axis cs:0.884,-0.634) {$R_{\text{max}}\text{ (km)}$};

\end{axis}

\begin{axis}[
    at={(2.8cm,5.25cm)},
    anchor=south,
    width=1cm,
    height=0.22cm,
    scale only axis,
    hide axis,
    xmin=0, xmax=1,
    ymin=0, ymax=1,
    colorbar horizontal,
    colormap={mycolormap}{
        color(0cm)=(mycolor1);
        color(0.25cm)=(mycolor2);
        color(0.5cm)=(mycolor3);
        color(0.75cm)=(mycolor4);
        color(1.0cm)=(mycolor5);
        color(1.25cm)=(mycolor6);
        color(1.5cm)=(mycolor7);
        color(1.75cm)=(mycolor8);
        color(2.0cm)=(mycolor9)
    },
    point meta min=2,
    point meta max=200,
    colorbar style={
        width=1.4cm,
        height=0.22cm,
        xtick={10, 100, 200},
        xticklabel style={font=\scriptsize},
        label style={font=\scriptsize},
        xlabel={$ \Delta f~(\text{kHz}) $},
        xlabel style={at={(0.5,-1.2)}, font=\scriptsize},
        xtick align=inside,
    }
]
\addplot [draw=none] coordinates {(0,0)};
\end{axis}

\end{tikzpicture}
    \vspace{-1.cm}
  \centerline{(a)} \medskip
\end{minipage}
\begin{minipage}[b]{0.49\linewidth}
  \centering
    \definecolor{mycolor1}{rgb}{0.00000,0.44700,0.74100}%
\definecolor{mycolor2}{rgb}{0.85000,0.32500,0.09800}%
\definecolor{mycolor3}{rgb}{0.92900,0.69400,0.12500}%

\begin{tikzpicture}
\begin{axis}[%
    width=3.2cm,
    height=5cm,
    scale only axis,
    xmin=2,
    xmax=16,
    xtick={4,8,12,16},
    xlabel={Antennas Per Satellite},
    ymin=0.2,
    ymax=1.4,
    ylabel={RMSE / $\Delta R$},
    xlabel style={font=\footnotesize},
    ylabel style={font=\footnotesize},
    tick label style={font=\scriptsize},
    xmajorgrids,
    ymajorgrids,
    axis background/.style={fill=white},
    axis x line*=bottom,
    axis y line*=left,
    legend style={
        font=\scriptsize,
        cells={anchor=west},
        at={(0.97,0.97)},
        anchor=north east,
        draw=black,
        fill=white,
        align=left
    }
]

\addplot [color=mycolor1, line width=2pt, mark size=4pt, mark=o, mark options={solid, mycolor1}]
table[row sep=crcr]{
2	1.2452\\
3	0.7947\\
4	0.6755\\
6	0.5551\\
8	0.5122\\
16	0.4611\\
};
\addlegendentry{Single-LEO}

\addplot [color=mycolor2, line width=2pt, mark size=2.8pt, mark=square, mark options={solid, mycolor2}]
table[row sep=crcr]{
2	1.2352\\
3	0.7938\\
4	0.6102\\
6	0.4282\\
8	0.3662\\
16	0.3251\\
};
\addlegendentry{Multi-LEO LEF}

\addplot [color=mycolor3, line width=2pt, mark size=2.7pt, mark=triangle, mark options={solid, mycolor3}]
table[row sep=crcr]{
2	0.379\\
3	0.3577\\
4	0.3526\\
6	0.3496\\
8	0.3489\\
16	0.3217\\
};
\addlegendentry{Multi-LEO DFE}

\end{axis}
\end{tikzpicture}
    \vspace{-1.cm}
  \centerline{(b)} \medskip
\end{minipage}
\vspace{-0.8cm}
\caption{(a) Maximum unambiguous range $R_{\text{max}}$, range resolution $\Delta R$, and maximum unambiguous velocity $v_{\text{max}}$ as functions of subband spacing ($\Delta f$ ranging from $1$ kHz to $200$ kHz).
(b) Comparison between single-\ac{leo} satellite monostatic sensing and multi-\ac{leo} satellite multistatic sensing with two strategies: LEF and DFE. To support an unambiguous range of $100~\mathrm{km}$ and a relative velocity of $7.5~\mathrm{km/s}$, the subband spacing and symbol duration are set to $1.5~\mathrm{kHz}$ and $1.5~\mu\mathrm{s}$, respectively, in (b). In radar systems, the symbol duration should be greater than or equal to the inverse of the subband spacing; for illustration purposes, equality is assumed in (a).}
\label{sensing_compare}
\vspace{-0.2cm}
\end{figure}
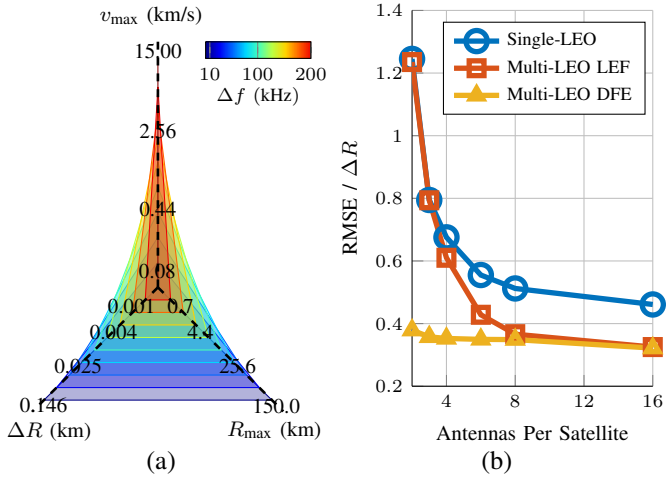

Figure~\ref{sensing_compare}(a) illustrates the trade-offs introduced by varying the subband spacing. Specifically, increasing the subband spacing improves the maximum unambiguous velocity but simultaneously decreases the maximum unambiguous range and range resolution. While larger subband spacings are favored in communication-centric \ac{ofdm} designs, they can lead to range aliasing and hinder long-range sensing performance. To maintain communication efficiency while improving sensing reliability, advanced algorithmic strategies must be considered. These include multi-snapshot processing, frequency ramping techniques (e.g., enabled by the Chinese remainder theorem), and joint estimation methods that exploit received signal strength. Consequently, waveform design, signal modeling, and estimation algorithms must be co-optimized to effectively balance the trade-offs between sensing accuracy and communication throughput in \ac{leo} satellite constellations.

We illustrate the advantages of multi-\ac{leo} satellite multistatic sensing. Fig. \ref{sensing_compare}(b) compares three approaches: (i) single-\ac{leo} satellite monostatic sensing, (ii) multi-\ac{leo} satellite multistatic sensing using local-estimate-then-fusion (LEF) across four satellites, which requires only time synchronization, and (iii) multistatic sensing with data-fusion-then-estimate (DFE), which demands stringent phase synchronization among satellites. Results show that LEF can achieve performance comparable to the more complex DFE strategy when each satellite collects a sufficient number of antenna observations, offering a more practical solution by relaxing synchronization requirements. Moreover, the spatial diversity inherent in multistatic configurations helps mitigate Doppler-induced distortions resulting from satellite-target relative motion, thereby improving sensing robustness in dynamic scenarios.

\section{\ac{dislac} Over \ac{leo} Satellite Constellation — A System-Level Perspective}
Unifying communication, localization, and sensing within a single spaceborne infrastructure, i.e., \ac{leo} satellite constellation, offers promising reductions in operational (OPEX) and, to a lesser extent, capital expenditures (CAPEX) when adressing the constellation launch. Building on insights from terrestrial 6G networks \cite{henk2025lens,henk2025dmimo}, this vision is extended here to the context of \ac{dislac} over \ac{leo} satellite constellations. However, the actual benefits of resource sharing across functionalities depend not only on waveform commonality (e.g., \ac{ofdm}) but, more critically, on their compatibility at the concept of operations level. The goal is not to revisit single-function systems such as direct-to-cell architectures \cite{tuzi2023access}, but to identify shared system-level enablers, spanning synchronization, \ac{isl} design, and onboard processing, that make \ac{dislac} over \ac{leo} constellations a viable and scalable solution. 
Key implications across these dimensions are summarized in Table~\ref{tab:SystemAspects}.

The case studies in Section~III have illustrated how multi-\ac{leo} cooperation influences each functionality individually. In a unified \ac{dislac} framework, these functionalities become mutually reinforcing. On the one hand, accurate sensing and localization strengthen communication through more reliable user association, seamless handover, and efficient resource management. On the other hand, advanced communication capabilities further enhance sensing and localization by enabling high-quality signal transmission, flexible information exchange, and greater spatial diversity. Section~IV therefore builds on the observations from Section~III and focuses on the \ac{ntn}-specific, system-level aspects of \ac{dislac} integration, such as synchronization, \acp{isl}, constellation topology, and regulatory constraints, rather than repeating numerical comparisons for a single joint experiment.

\begin{table*}[!t]
\centering
\caption{System-Level Considerations for DISLAC over LEO Satellite Constellations}
\label{tab:SystemAspects}
\renewcommand{\arraystretch}{1.15}
\begin{tabular}{p{0.12\textwidth} p{0.18\textwidth} p{0.18\textwidth} p{0.18\textwidth} p{0.22\textwidth}}
\toprule
\textbf{Aspect} & 
\textbf{Communication} & 
\textbf{Localization} & 
\textbf{Sensing} & 
\textbf{DISLAC Proposal} \\
\midrule
\textbf{Synchronization \& Coherency} &
\cellcolor{cyan!5} 
\textbf{\textit{Requirement:}} 10--100\,ns for macro-diversity (TSN);    $<1$\,ns for coherent beamforming (PSN).\par
\textbf{\textit{Challenge:}} LEO oscillator drift; GNSS alone insufficient (PSN); needs TWTFT and on-the-fly calibration of Tx chain(s). &
\cellcolor{green!5}
\textbf{\textit{Req.:}} $\sim$10\,ns for delay/Doppler; $<1$\,ns for carrier phase.\par
\textbf{\textit{Chal.:}} Strong orbit/Doppler variations; GNSS degradation;  satellite selection w.r.t. local clock stability to support tight synchronization. &
\cellcolor{blue!5}
\textbf{\textit{Req.:}} $<1$\,ns for coherent multistatic (PSN); 10--100\,ns for non-coherent (TSN).\par
\textbf{\textit{Chal.:}} Drift, beam-steering phase shifts, high Doppler. &
\cellcolor{red!5}
Hybrid TSN/PSN: intra-cluster PSN with sub-ns synchronization via TWTFT \acp{isl}; inter-cluster TSN via GNSS; adaptive satellite selection by synchronization quality. \\
\midrule
\textbf{Tx/Rx mode \& Antennas} &
\cellcolor{cyan!5} 
\textbf{\textit{Req.:}} Typically single Tx/Rx mode per carrier frequency; multi-beam; steerable for high platform dynamic.\par
\textbf{\textit{Chal.:}} Limited RF chains; fixed vs steerable trade-off. &
\cellcolor{green!5}
\textbf{\textit{Req.:}} Beamformed Tx or orthogonal PRS  Tx.\par
\textbf{\textit{Chal.:}} Antenna sharing; limited simultaneous beams; PRS scheduling. &
\cellcolor{blue!5}
\textbf{\textit{Req.:}} Full-Duplex (monostatic) or asymmetric Tx/Rx (multistatic); steerable beams.\par
\textbf{\textit{Chal.:}} Duplex mismatch; isolation; agility limits. &
\cellcolor{red!5}
Reconfigurable hybrid arrays; full-duplex on sensing nodes; dynamic beam orchestration to guarantee communication/localization/sensing coverage. \\
\midrule
\textbf{\acp{isl} \& Topology} &
\cellcolor{cyan!5} 
\textbf{\textit{Req.:}} \acp{isl} for traffic offload in TSNs; low-latency beamforming coordination and tight synchronization in PSNs.\par
\textbf{\textit{Chal.:}} Dynamic topology; optical pointing errors. &
\cellcolor{green!5}
\textbf{\textit{Req.:}} \acp{isl} for synchronization, ephemerides generation and distribution, PRS configuration.\par
\textbf{\textit{Chal.:}} Frequent neighbors/topology changes; bandwidth limits. &
\cellcolor{blue!5}
\textbf{\textit{Req.:}} \acp{isl} for metadata/raw echoes (DFE).\par
\textbf{\textit{Chal.:}} High data throughput peaks; synchronization/metadata overhead. &
\cellcolor{red!5}
Multi-tier \acp{isl}: \ac{rf} for synchronization/control, optical for high-rate; adaptive Ring $\leftrightarrow$ Star switching; selective raw-data transfer for key sensing events. \\
\midrule
\textbf{Standards \& Regulations} &
\cellcolor{cyan!5} 
\textbf{\textit{Req.:}} Compliance to MSS compatibility; NTN coexistence.\par
\textbf{\textit{Chal.:}} Multi-LEO collaborative beamforming in 3GPP; cross-border spectrum. &
\cellcolor{green!5}
\textbf{\textit{Req.:}} PRS vs. NTN traffic compliance; interoperability with terrestrial networks.\par
\textbf{\textit{Chal.:}} Provision of PNT service with performance beyond pure communication needs MSS protected band. &
\cellcolor{blue!5}
\textbf{\textit{Req.:}} Radar spectrum/licensing; debris mitigation.\par
\textbf{\textit{Chal.:}} Sharing with communication/localization services; constellation size limits. &
\cellcolor{red!5}
Harmonized spectrum policy with dynamic sub-band allocation; \ac{dislac}-specific 3GPP/ITU signaling; integrated debris-mitigation plan. \\
\bottomrule
\end{tabular}
\end{table*}

\subsection{Synchronization, Positioning, and Signal Coherency}

Distributed satellite architectures can be broadly classified as either time-synchronized networks (TSN) or phase-synchronized networks (PSN), each imposing distinct requirements on system design and retained technologies. TSNs rely on a shared time reference to align transmission and reception across \ac{leo} satellites, enabling functionalities such as traffic rebalancing and spatial diversity for communication, localization, and sensing. In contrast, PSNs aim to coherently combine signals across satellites to boost received power, demanding far tighter phase-level synchronization and often requiring formation flying (e.g., swarms) to maintain signal alignment \cite{tuzi2023access}. The choice between TSN and PSN directly impacts the required timing and positioning accuracy. In TSNs, meter-level accuracy and synchronization between 10 to 100ns (as provided by \ac{gnss} and \ac{3gpp} SIB19) are generally sufficient for communication. However, in PSNs and for high-precision localization and target sensing, tighter accuracies which are strongly frequency dependent, becomes essential. While \ac{gnss}-based solutions can provide a coupled position-time reference, alternatives such as two-way time and frequency transfer (TWTFT) over \acp{isl} allow for decoupling and enhanced flexibility\cite{Liz2022access}. Relative positioning suffices for intra-cluster operations, while absolute positioning is required for inter-cluster communication and coordination. 

Moreover, tight time and phase synchronization alone is not sufficient for coherent distributed arrays. Ensuring signal phase coherency, particularly at higher carrier frequencies, requires mechanisms for on-the-fly calibration of the transmission chain group delays. These may involve monitoring nodes within the swarm or external ground-based references, similar to established \ac{sar} calibration techniques. Steerable beams, while beneficial for flexibility, introduce additional complexity and cost to the calibration process. Beyond phase alignment, signal distortion must also be minimized, as asymmetries in the correlation or cross-ambiguity functions degrade localization and sensing accuracy.

\subsection{Tx/Rx Mode and Antenna Flexibility}

In \ac{leo} satellite constellations, the three core functionalities, communication, localization, and sensing, impose distinct requirements on transmission/reception modes in a duplex scheme and antenna flexibility. Communication and localization typically operate in either transmit or receive mode at distinct carrier frequencies (cases of the Globalstar and more recent broadband Starlink, OneWeb and Kuiper systems), while sensing obviously necessitates transmission and reception at same carrier frequency. Although basic radar systems alternate between transmit and receive phases, advanced implementations exploit MIMO and polarization techniques to achieve concurrent operation. These differences pose challenges to achieving full payload-level synergy, particularly between sensing and the two other functions. One approach to resolving these conflicts is time-division multiplexing of communication, localization, and sensing sessions, though this compromises service availability. Alternatively, multistatic radar configurations with asymmetric transmit/receive assignments across or within satellite clusters offer more flexible solutions. However, sensing generally demands additional payload complexity and satellite resources beyond what communication and localization require.

Antennas are central shared resources in this context and must accommodate the operational needs of all three functionalities. A key consideration is the number of beams that can be simultaneously formed for transmission or reception, enabling spatial multiplexing and frequency reuse, essential for communications and also critical for \ac{prs}-based localization per \ac{3gpp} standards. In contrast, sensing may require only a single beam for space target tracking, unless simultaneous and multiple targets tracking is desired. Another important factor is beam steering. \ac{3gpp} distinguishes between fixed beams (for moving tracking areas) and steerable beams (for fixed tracking areas), which influence service strategies and hardware complexity. Similar principles apply in sensing, where scanning modes use fixed beams, while spotlight or tracking modes rely on steerable beams.

\subsection{ISL and Constellation Topology}

\acp{isl} are foundational to large-scale \ac{leo} constellations, reducing reliance on ground infrastructure and minimizing latency. In \ac{dislac}, their role becomes even more critical for enabling continuous inter-satellite coordination. TSN architectures, which tolerate larger inter-satellite distances, can adopt conventional \ac{rf} or optical \acp{isl}, though this increases scheduling complexity. PSN architectures, by contrast, may require dual \acp{isl} payloads, or even different technologies, for intra- and inter-swarm links, to support more stringent synchronization. Intra-swarm \acp{isl} must accommodate mission traffic data and auxiliary data such as collaborative beamforming parameters (as in Case Study 1), relative positions, and timing updates. Integrated inter-satellite ranging further supports precise synchronization and positioning. While communication imposes the highest \ac{isl} throughput demands, sensing and localization typically require less bandwidth. However, exceptions exist, for instance, in multistatic sensing where full-frame correlation is employed to enhance resolution, substantially increasing \ac{isl} data burden. 

These requirements are closely tied to \ac{leo} satellite constellation topology, which must be harmonized across the three functionalities to balance performance with cost, deployment/replenishment logistics, and regulatory compliance. While legacy communication typically requires one or two links per user, localization and sensing benefit from greater spatial diversity, requiring multiple satellites in view. Even communication can demand enhanced visibility in TSNs for traffic load balancing. A practical \ac{dislac} system should therefore aim to maximize the number of simultaneously visible and exploitable satellites.

\subsection{Standardizations and Regulations}
The practical deployment of \ac{dislac} systems must comply with existing and emerging standardization and regulatory frameworks, ranging from spectrum management to orbital sustainability. In particular, the legacy approach asking each communication, navigation or sensing system operator to comply with the regulations established at International Telecommunication Union (ITU) for a mobile satellite service (MSS), radio navigation-satellite service (RNSS) or Earth exploration satellite service (EESS), needs to be addressed when integrating all three services in a single system. At first glance, a joint compliance over all services could further constrain the \ac{dislac} payload and network design. However, \ac{dislac} also represents an opportunity to alleviate such burden, building on the exploitation of multi-purpose signals, which revisits the notion of primary or secondary services (i.e., co-existing in the same band) depending on the use of those signals. Furthermore, realizing inter-satellite cooperation at scale will require extensions to current 3GPP \ac{ntn} specifications to support standardized inter-satellite signaling, synchronization, and multi-function coexistence. Beyond spectrum policy, space sustainability and debris-mitigation regulations impose additional constraints on constellation size, orbit selection, and resource reuse, urging more efficient and environmentally responsible deployment strategies. By integrating these technical and regulatory considerations, \ac{dislac} provides a forward-looking framework that links system-level innovation with the industrial standardization pathways essential for large-scale multi-satellite collaboration.

\section{Emerging Challenges and Research Directions}

In addition to the system-level considerations summarized in Table~\ref{tab:SystemAspects}, several additional challenges must be addressed to realize the full benefits of \ac{dislac} over \ac{leo} satellite constellations. The following subsections discuss key research directions from a \ac{dislac} perspective. 

\subsection{Collaborative Beamforming over Stochastic \ac{leo} Satellite Constellation Topology} 
Unlike terrestrial networks, where access points are typically fixed with rigid topology, \ac{leo} satellite constellations exhibit a dynamic topology that evolves over time due to orbital motion. Moreover, instead of relying on stable fiber-wired backhaul as in terrestrial systems, \ac{leo} satellites are interconnected via \acp{isl}, often implemented using highly directional links (e.g., optical), which are prone to pointing errors and link interruptions. These features render \ac{leo} constellations inherently stochastic, making conventional collaborative beamforming schemes, developed for deterministic terrestrial networks, unsuitable for direct application. 
Future work should therefore explore topology-aware and synchronization-aided beamforming strategies that jointly exploit localization and sensing information to improve link alignment, mitigate Doppler uncertainty, and enhance multi-\ac{leo} coordination robustness within the \ac{dislac} framework. 

\subsection{Satellite Selection and Orbit Uncertainties}
An important research direction lies in improving localization performance by selectively incorporating or excluding satellite observations based on their time/phase synchronization and state-estimation quality. Not all \ac{leo} satellites may maintain accurate knowledge of their own positions and velocities or remain tightly synchronized within the constellation, resulting in imperfect timing advance and Doppler pre-compensation. In such cases, blindly aggregating data from all available satellites can degrade localization accuracy. A more nuanced approach is to assess each satellite’s reliability in real time and weigh its contribution accordingly. 
Orbit and state uncertainties, inherent to \ac{leo} constellations due to frequent maneuvers and limited onboard calibration, can also propagate to sensing and communication functions through synchronization and beam alignment errors. 
Hence, developing joint reliability metrics and cross-functional calibration methods will be essential to ensure robust and mutually consistent localization, sensing, and communication performance across the constellation. 

\subsection{Constellation Optimization and Integration with Terrestrial Systems}
The performance of distributed \ac{leo}-based sensing and localization strongly depends on constellation geometry and its integration with terrestrial infrastructure. For multi-satellite sensing, estimation accuracy, characterized by the \ac{crb}, is critically influenced by satellite positions. Similarly, wider spatial apertures improve localization accuracy by reducing geometric dilution of precision (GDOP), yet also introduce larger signal delays and Doppler spreads, complicating synchronization and waveform reuse across communication and sensing functions. These trade-offs necessitate constellation designs that adapt geometry and resource allocation according to joint service requirements. 
Maneuverable modern \ac{leo} satellites can dynamically adjust their geometry to balance wide-area coverage and focused beamforming, enabling task-driven reconfiguration for both data and sensing missions. 
Furthermore, integrating \ac{leo}-based systems with terrestrial \ac{dmimo} and reconfigurable intelligent surfaces (RISs) offers additional cross-domain diversity and resource sharing. 
Such integration demands flexible architectures capable of fusing multi-source information for global-scale \ac{dislac} operation. 

\subsection{New Waveforms for Space Sensing with Communication Compatibility} 
Waveforms critically influence sensing and communication performance through characteristics such as low sidelobes, Doppler resilience, and spectral efficiency. As discussed in Section~\ref{ssec:CS2}, robust waveform behavior under large Doppler shifts is essential in \ac{leo} environments. 
Although pre- and post-compensation techniques are commonly employed, their effectiveness remains limited under fast-varying dynamics. Frequency-modulated continuous-wave (FMCW) waveforms offer strong Doppler tolerance and simplified receivers but cannot efficiently support high data rates. Conversely, \ac{ofdm} waveforms provide high throughput yet lose subcarrier orthogonality under large Doppler spreads, degrading both communication and sensing performance. 
These contrasting properties motivate the design of \emph{communication-compatible} sensing waveforms that balance data rate, ambiguity resolution, and Doppler tolerance, such as \ac{otfs}-based designs. 
Developing adaptive waveform strategies that jointly optimize communication throughput, localization precision, and sensing accuracy will be key to fully realizing the potential of \ac{dislac} systems.

\section{Acknowledgments}
This work is supported by KAUST Office of Sponsored Research (OSR) under Award No. RFS-CRG12-2024-6478 and Global Fellowship Program under Award No. RFS-2025-6844.
The views expressed herein are those of the authors and do not necessarily reflect the official opinion of Airbus. 

\vspace{-1cm}
\begin{IEEEbiographynophoto}{Yuchen~Zhang}
(yuchen.zhang@kaust.edu.sa) is a Postdoctoral Global Fellow in the Electrical and Computer Engineering Program, King Abdullah University of Science and Technology (KAUST), Kingdom of Saudi Arabia. His current research focuses on 6G \ac{ntn} communication and positioning.
\end{IEEEbiographynophoto}

\vspace{-1cm}
\begin{IEEEbiographynophoto}{Francis Soualle}
(francis.soualle@airbus.com) is an expert and senior R\&D navigation engineer at Airbus Defence and Space, Germany. His main focuses cover architectural trades, design and performance of satellite-based PNT systems. His current research interests include fused COM-NAV solutions building on 5G and 6G systems.
\end{IEEEbiographynophoto}

\vspace{-.8cm}
\begin{IEEEbiographynophoto}{Musa Furkan Keskin}
(furkan@chalmers.se) is a Researcher at Chalmers University of Technology, Gothenburg, Sweden. His current research interests include integrated sensing and communications, and hardware impairments in beyond 5G/6G systems. 
\end{IEEEbiographynophoto}

\vspace{-1cm}
\begin{IEEEbiographynophoto}{Yuan~Liu}
(yuan.liu@uni.lu) is a Postdoctoral Researcher at the Interdisciplinary Centre for Security, Reliability and Trust (SnT), University of Luxembourg, L-1855, Luxembourg. His current research focuses on channel modeling and signal processing for active sensing applications.
\end{IEEEbiographynophoto}

\vspace{-1cm}
\begin{IEEEbiographynophoto}{Linlong~Wu}
(linlong.wu@uestc.edu.cn) is a professor with the School of Information and Communication Engineering, University of Electronic Science and Technology of China (UESTC), Chengdu, China. His research focuses on signal processing, radar, ISAC and RIS.
\end{IEEEbiographynophoto}

\vspace{-1cm}
\begin{IEEEbiographynophoto}{José~A.~del~Peral-Rosado}
(jose\_antonio.del\_peral\_rosado@airbus.com) is a senior R\&D navigation engineer at Airbus Defence and Space, Germany. His current research focuses on multi-layer PNT systems and demonstrators with GNSS, 5G and 6G.
\end{IEEEbiographynophoto}

\vspace{-1cm}
\begin{IEEEbiographynophoto}{Bhavani Shankar M.~R}
(Bhavani.Shankar@uni.lu) is an Assistant Professor at SnT, University of Luxembourg and his research interests include radar signal processing and wireless communications. 
\end{IEEEbiographynophoto}

\vspace{-1cm}
\begin{IEEEbiographynophoto}{Gonzalo~Seco-Granados}
(gonzalo.seco@uab.cat) is a professor with the Department of Telecommunication, Universitat Autònoma de Barcelona, Bellaterra, Spain. His research interests include GNSS, and beyond 5G integrated communications, localization, and sensing.
\end{IEEEbiographynophoto}

\vspace{-1cm}
\begin{IEEEbiographynophoto}{Henk Wymeersch}
(henkw@chalmers.se) is a professor at the Electrical Engineering Department, Chalmers University of Technology, Sweden. His current research focuses radio localization and sensing for 5G and 6G.
\end{IEEEbiographynophoto}

\vspace{-1cm}
\begin{IEEEbiographynophoto}{Tareq~Y.~Al-Naffouri}
(tareq.alnaffouri@kaust.edu.sa) is a professor in the Electrical and Computer Engineering Program, King Abdullah University of Science and Technology (KAUST), Kingdom of Saudi Arabia. His current research focuses on statistical inference and learning, with applications in wireless communications, positioning, navigation, smart cities, and smart health.
\end{IEEEbiographynophoto}

\bibliography{references}
\bibliographystyle{IEEEtran}

\end{document}